\documentclass[prd,aps,preprint,nofootinbib,superscriptaddress]{revtex4}
\usepackage{epsfig}
\usepackage{amsmath}

\begin{document}

\preprint{BNL-NT-06/15} \preprint{RBRC-598}

\title{Single-Transverse Spin Asymmetry in Semi-Inclusive \\
Deep Inelastic Scattering}

\author{Xiangdong Ji}
\email{xji@physics.umd.edu} \affiliation{Physics Department,
University of Maryland, College Park, MD 20742}
\author{Jian-Wei Qiu}
\email{jwq@iastate.edu} \affiliation{Department of Physics and
Astronomy, Iowa State University, Ames, IA 50011}
\author{Werner Vogelsang}
\email{vogelsan@quark.phy.bnl.gov} \affiliation{Physics
Department, Brookhaven National Laboratory, Upton, NY 11973}
\affiliation{RIKEN BNL Research Center, Building 510A, Brookhaven
National Laboratory, Upton, NY 11973}
\author{Feng Yuan}
\email{fyuan@quark.phy.bnl.gov} \affiliation{RIKEN BNL Research
Center, Building 510A, Brookhaven National Laboratory, Upton, NY
11973}
\date{\today}

\begin{abstract}
We study the single-transverse spin asymmetry in semi-inclusive
hadron production in deep inelastic scattering. We derive the
leading contribution to the asymmetry at moderate transverse
momentum $P_{h\perp}$ of the produced hadron in terms of
twist-three quark-gluon correlation functions, and compare with
the approach based on the factorization at fixed transverse
momentum involving the asymmetric transverse-momentum and
spin-dependent quark distribution. We verify that the two
approaches yield identical results in this regime. By a comparison
with our earlier calculations for the single-spin asymmetry in the
Drell-Yan process we recover the well-established
process-dependence of the time-reversal-odd
transverse-momentum-dependent quark distributions that generate
single-spin phenomena.
\end{abstract}

\maketitle

\newcommand{\be}{\begin{equation}}
\newcommand{\ee}{\end{equation}}
\newcommand{\ben}{\[}
\newcommand{\een}{\]}
\newcommand{\beqn}{\begin{eqnarray}}
\newcommand{\eeqn}{\end{eqnarray}}
\newcommand{\Tr}{{\rm Tr} }

{\bf 1.} The study of single-transverse spin asymmetries
(SSAs) has been at the forefront of experimental and theoretical
research in strong interaction physics ever since the first observation
of strikingly large asymmetries in hadronic scattering in the
1970s~\cite{E704-Bunce}. The size of the asymmetries posed
a significant challenge for QCD. With the advent of new experimental
information from lepton scattering~\cite{dis} and from RHIC~\cite{rhic},
and with major recent theory advances, we are now beginning to obtain a much
clearer picture of the possible origins of SSAs in QCD~\cite{review}.

In particular, two mechanisms for generating SSAs had been
identified in the literature: asymmetric
transverse-momentum-dependent parton distribution in a
transversely polarized proton (the so-called Sivers functions)
\cite{Siv90}, and twist-three transverse-spin-dependent
quark-gluon correlation functions (the so-called
Efremov-Teryaev-Qiu-Sterman (ETQS) mechanism) \cite{Efremov,qiu}.
For a long time, despite a wide-spread belief that these two
mechanisms were not completely unrelated, the precise connection
between them remained obscure. Early efforts to link the two were
made in~\cite{BoeMulPij03,MaWa03,bacchetta}. In two recent
publications \cite{JiQiuVogYua06}, we have demonstrated that the
two mechanisms each have their own domain of validity, and that
they consistently describe the same physics in kinematic regime
where they both apply. We have shown this in \cite{JiQiuVogYua06}
for the case of the SSA for Drell-Yan production of dilepton pairs
with invariant mass $Q$ and transverse momentum $q_\perp$. At
large $q_\perp\sim Q$, the ETQS mechanism applies, and the
resulting SSA is of twist-three nature. At small $q_\perp\ll Q$, a
factorization in terms of transverse-momentum dependent (TMD)
parton distribution
applies~\cite{ColSop81,ColSopSte85,JiMaYu04,ColMet04}, involving
in case of the SSA the Sivers functions. If $q_\perp$ is much
larger than $\Lambda_{\rm QCD}$, the dependence of these functions
on transverse momentum may be computed using QCD perturbation
theory. At the same time, the result obtained within the ETQS
formalism may also be extrapolated into the regime $\Lambda_{\rm
QCD}\ll q_\perp \ll Q$, and we demonstrated in
Ref.~\cite{JiQiuVogYua06} that the result of this extrapolation is
identical to that obtained using the TMD approach. In this sense,
we have unified the two mechanisms widely held responsible for the
observed SSAs.

In the present letter, we extend our previous calculations to the
case of semi-inclusive hadron production in lepton-hadron deep
inelastic scattering (SIDIS) \cite{sidis}, $e (\ell)+p(P)\to
e(\ell') + h (P_h) + X$, which proceeds through exchange of a
virtual photon with momentum $q_\mu=\ell_\mu-\ell'_\mu$ and
invariant mass $Q^2=-q^2$. A similar calculation has also been
reported in \cite{yujitalk} in the quark-gluon correlation
approach, where the twist-three effects in the parton distribution
\cite{qiu} as well as in the fragmentation function
\cite{Kanazawa:2000hz} were considered, and they focused on the
SSAs at large transverse momenta, $P_{h\perp}\sim Q$, with
$P_{h\perp}$ the transverse momentum of the final-state hadron in
the ``hadron frame'' defined below. Our present calculation
extends the work of~\cite{yujitalk}. In particular, only the
so-called ``derivative'' contributions were taken into account
in~\cite{yujitalk}. These may or may not dominate the
spin-dependent cross section at large $P_{h\perp}\sim Q$. When
$P_{h\perp}\ll Q$, there are definitely other equally important
contributions, which we will calculate. More importantly, we are
interested in hadron production at intermediate transverse
momenta, $\Lambda_{\rm QCD}\ll P_{h\perp}\ll Q$, where we will
compare the predictions from the two mechanisms. At first sight,
this additional verification of the consistency of the two
mechanisms might appear to be of rather limited interest. However,
there are several reasons why we believe that this is a valuable
addition. Foremost, the SIDIS process is of greater current
interest experimentally than Drell-Yan, with several experiments
producing data for SSAs in lepton scattering~\cite{dis}. We stress
that, apart from clarifying the theoretical description of SSAs,
our work also provides a detailed scheme for the practical
analysis of single-spin asymmetries, since it addresses the
asymmetries over the whole kinematic regime of transverse
momentum. Secondly, as is well-known by now, the peculiar
gauge-dependence properties of the Sivers
functions~\cite{BroHwaSch02,Col02,BelJiYua02} predict a sign
change of the functions when going from the Drell-Yan process to
SIDIS. It is important to verify this sign change in an explicit
calculation of a physical process, and our way of doing this is to
confront our earlier Drell-Yan calculation with that for SIDIS.
This provides a test of the QCD factorization and of the
(non-)universality of spin-dependent TMD parton distributions.

The presentation of this paper will very closely follow our
previous work. We will start by calculating the SSA for SIDIS at
large transverse momentum of the produced hadron, $P_{h\perp}\sim
Q$. We will then expand the obtained result for $P_{h\perp}\ll Q$,
in order to make contact with the expression provided by TMD
factorization~\cite{JiMaYu04}, and we will verify that also for
SIDIS both approaches contain the same physics in the region
$\Lambda_{\rm QCD}\ll P_{h\perp}\ll Q$. For this to hold true, the
sign change mentioned above is vital. In this paper, we focus
entirely on the single-spin asymmetries coming from the
quark-gluon correlation function and/or the Sivers functions in
the polarized proton (referred to as the ``Sivers-type" SSA in the
following) in SIDIS. There are also other contributions to the
SSA, for example associated with the quark-gluon correlation in
the fragmentation functions and the so-called Collins
effect~\cite{Col93,yujitalk}. We reserve the study of these for
future work.

{\bf 2.} We start by calculating the single-transverse-spin
dependent differential cross section for SIDIS at large transverse
momentum $P_{h\perp}$ of the produced hadron (defined below in the
``hadron frame''), where the ETQS formalism is relevant. The
differential cross section may be calculated from the formula
\begin{equation}
    \frac{d\sigma(S_\perp)}{dx_Bdydz_hd^2\vec{P}_{h\perp}}
      = \frac{2\pi\alpha^2_{\rm em}}{Q^4}
      yL_{\mu\nu}(\ell,q)W^{\mu\nu}(P, S_\perp,q, P_h) \ ,
\end{equation}
where $\alpha_{\rm em}$ is the electromagnetic coupling and
$x_B\equiv Q^2/2 P\cdot q$, $z_h \equiv P\cdot P_h/P\cdot q$,
$y\equiv P\cdot q/P\cdot \ell$. We also introduce
$S_{ep}=(P+\ell)^2$, the electron-proton center of mass energy
squared. $L^{\mu\nu}$ and $W_{\mu\nu}$ are the leptonic and
hadronic tensors, respectively. The latter depends on the
transverse proton spin vector, $S_\perp$. We consider scattering
of unpolarized leptons by virtual-photon exchange, in which case
the leptonic tensor is given by
\begin{eqnarray}
L^{\mu\nu}(\ell,q) &=& 2\left(\ell^\mu{\ell'}^\nu + \ell^\mu{\ell'}^\nu -
g^{\mu\nu}Q^2/2\right) \ .
\end{eqnarray}
The hadronic tensor has the following expression in QCD:
\begin{equation}
   W^{\mu\nu}(P, S_\perp, q, P_h) =\frac{1}{4z_h}\sum_X
   \int \frac{d^4\xi}{(2\pi)^4} e^{iq\cdot \xi}
   \langle PS|J_\mu(\xi)|XP_h\rangle \langle XP_h|J_\nu(0)|PS\rangle
   \ ,
\label{wmndef}
\end{equation}
where $J^\mu$ is the quark electromagnetic current and $X$
represents all other final-state hadrons other than the observed
particle $h$.

It is convenient to write the momentum of the virtual photon in
terms of the incoming and outgoing hadron momenta in SIDIS,
\begin{equation}
\label{qtdef}
q^\mu=q_t^\mu+\frac{q\cdot P_h}{P\cdot P_h}P^\mu+\frac{q\cdot
P}{P\cdot P_h}P_h^\mu \ ,
\end{equation}
with $q_t^\mu$ transverse to the momenta of the initial and final
hadrons, $q_t^\mu P_\mu = q_t^\mu P_{h\mu} = 0$. 
$q_t$ is a space-like vector; we define
\begin{equation}
\vec{q}_\perp^{\;2} \equiv - q_t^2
 = Q^2\left[1+\frac{1}{x_B}\,
               \frac{q\cdot P_h}{P\cdot P_h}\right]\ .
\label{qperpdef}
\end{equation}
The hadronic tensor $W^{\mu\nu}$ in Eq.~(\ref{wmndef})
can be decomposed in terms of five
parity and current conserving tensors
${\cal V}_i^{\mu\nu}$~\cite{sidis}:
\begin{equation}
W^{\mu\nu}=\sum\limits_{i=1}^5{\cal V}_i^{\mu\nu}W_i \ ,
\label{wdecomp}
\end{equation}
where the $W_i$ are structure functions which may be projected out
from $W^{\mu\nu}$ by
$W_i=W_{\alpha\beta}\tilde{\cal V}_i^{\alpha\beta}$,
with the corresponding inverse tensors $\tilde{\cal V}_i$
\cite{sidis}.  Both ${\cal V}_i$ and $\tilde{\cal V}_i$ can be
constructed from four orthonormal basis vectors~\cite{sidis}:
\begin{eqnarray}
T^\mu&=&\frac{1}{Q}\left(q^\mu+2x_B P^\mu\right) \ ,\nonumber\\
X^\mu&=&\frac{1}{q_\perp}\left[\frac{P_h^\mu}{z_h}-q^\mu-
\left(1+\frac{q_\perp^2}{Q^2}\right)x_B
P^\mu\right]\ ,\nonumber\\
Y^\mu&=&\epsilon^{\mu\nu\rho\sigma}Z_\nu X_\rho T_\sigma \ ,\nonumber\\
Z^\mu&=&-\frac{q^\mu}{Q} \ ,
\label{vecs}
\end{eqnarray}
with $q_\perp \equiv \sqrt{\vec{q}_\perp^{\; 2}}$ and
normalizations $T^\mu T_\mu=1, X^\mu X_\mu=Y^\mu Y_\mu=Z^\mu
Z_\mu=-1$. In the following, we will only consider the
contributions associated with the tensor ${\cal V}_1$. While the
tensor ${\cal V}_5$ does not contribute when contracted with a
symmetric $L_{\mu\nu}$, the other three tensors make contributions
that may also be relevant at large transverse momentum
$P_{h\perp}\sim Q$, and which need to be included in
phenomenological analyses~\cite{yujitalk}. However, as we
discussed in the Introduction, we are primarily interested in this
paper in hadron production in an intermediate transverse momentum
region, $\Lambda_{\rm QCD} \ll P_{h\perp} \ll Q$, and in the
connection between the ETQS mechanism and the TMD factorization
approaches, and we therefore want to investigate the limit
$P_{h\perp}\ll Q$ of the ETQS result. In that limit, ${\cal V}_1$
alone provides the leading behavior. This is known from the
literature~\cite{sidis} for the unpolarized cross section, and we
have verified it by explicit calculation for the (Sivers-type)
single-transverse-spin dependent polarized cross section. The
tensors ${\cal V}_1$ and $\tilde {\cal V}_1$ are given
by~\cite{sidis}
\begin{eqnarray}
{\cal V}_1^{\mu\nu}&=X^\mu X^\nu+Y^\mu Y^\nu, ~~~&\tilde{\cal
V}_1^{\mu\nu}=\frac{1}{2}\left(2T^\mu T^\nu+X^\mu X^\nu+Y^\mu
Y^\nu\right) \label{vdef} \ .
\end{eqnarray}

The definitions~(\ref{vecs}) for the coordinate vectors still leave
freedom to associate the axes with specific momentum directions. In the
following, we will perform our calculations in the so-called {\it hadron
frame}, where the virtual photon and target proton are taken to have
a spatial component only in the $z$-direction~\cite{sidis}:
\begin{equation}
P^\mu=P^+ p^\mu \ , ~~~~q^\mu=-x_BP^+p^\mu+\frac{Q^2}{2x_BP^+}n^\mu \ ,
\end{equation}
where the light-cone momenta are defined as $P^\pm=(P^0\pm
P^3)/\sqrt{2}$, and $p^\mu=(1^+,0^-,0_\perp)$, $n^\mu=(0^+,1^-,0_\perp)$
are two light-like vectors with $p\cdot n=1$. Usually one chooses
the photon to have a vanishing energy component, corresponding to
$P^+=Q/\sqrt{2}x_B$. In the hadron frame, the final state hadron will have the
momentum
\begin{equation}
P_h^\mu=\frac{x_B\vec{P}_{h\perp}^2}{z_h
Q^2}P^+p^\mu+z_h\frac{Q^2}{2x_BP^+}n^\mu+P_{h\perp}^\mu \ ,
\label{phdef}
\end{equation}
where $z_h$ has been defined above. Using the expression for
$q^\mu$ in~(\ref{qtdef}), one can show that in this frame $q_\perp
=P_{h\perp}/z_h$ with $P_{h\perp}=\sqrt{\vec{P}_{h\perp}^2}$. The
differential unpolarized and single-transverse-spin dependent
cross sections will be calculated in terms of $q_\perp$, which
will immediately give their dependence on $P_{h\perp}$. In the
following, we will use both $q_\perp$ and $P_{h\perp}$ when
discussing the transverse momentum in SIDIS, keeping in mind that
they are essentially the same in the hadron frame.

Substituting the tensors in Eq.~(\ref{vdef}) into~(\ref{wdecomp}) and
into the formula~(1) for the differential cross section, we obtain
\begin{eqnarray}
    \frac{d\sigma(S_\perp)}{dx_Bdydz_hd^2\vec{P}_{h\perp}}
      = \frac{4\pi\alpha^2_{\rm
      em}S_{ep}}{Q^4}\left\{2x_B(1-y+y^2/2)\, W_1      \right\}\ .
\end{eqnarray}

At large transverse momentum of the final state hadron, we may use
collinear factorization and compute $W_1$ in terms of parton
distribution functions, fragmentation functions for the produced
hadron, and hard partonic cross sections that may be calculated
using QCD perturbation theory. The lowest-order (LO) contributions
to the latter arise from the processes $\gamma^* q\to qg$ and
$\gamma^* g\to q\bar{q}$. Again, as we will eventually be interested
in the extrapolation of our results to $P_{h\perp}\ll Q$, we
focus on the part that will dominate at small $P_{h\perp}$, namely
the channel $\gamma^* q\to qg$. We then find for the
unpolarized SIDIS cross section:
\begin{eqnarray} \label{unp}
    \frac{d\sigma}{dx_Bdydz_hd^2\vec{P}_{h\perp}}\Big |_{{\cal V}_1}
      &=& \frac{4\pi\alpha^2_{\rm
      em}S_{ep}}{z_h^2Q^4}\frac{\alpha_s}{2\pi^2}C_F
\int \frac{dxdz}{xz}q(x)\hat q
      (z)\nonumber\\
      &&\times x_B\left(1-y+\frac{y^2}{2}\right)\hat \sigma_{unp}
      ~\delta\left(\vec{q}_\perp^{\;2}-\frac{Q^2(1-\xi)(1-\hat \xi)}
{\xi\hat \xi}\right)\ ,
\end{eqnarray}
where
the contribution $\hat\sigma_{unp}$ associated with the tensor structure
${\cal V}_1$ has been given in the literature~\cite{sidis}:
\begin{eqnarray}
      \hat \sigma_{unp}=\xi \hat\xi
      \left[\frac{1}{Q^2\vec{q}_\perp^{\;2}}\left(\frac{Q^4}{\xi^2\hat
      \xi^2}+(Q^2-\vec{q}_\perp^{\;2})^2\right)+6\right] \
      .\label{sunp}
\end{eqnarray}
Here the variables $\xi$ and $\hat \xi$ are defined as $\xi=x_B/x$ and
$\hat\xi=z_h/z$, with $x$ and $z$ the initial- and final-state partonic
momentum fractions, respectively. $q(x)$ denotes the quark parton
distribution function, and $\hat q(z)$ the fragmentation function
for a quark going into the observed hadron. We have for simplicity
suppressed their dependence on a factorization scale, and also a sum over
all quark and anti-quark flavors in~(\ref{unp}). Finally, $\alpha_s$
denotes the strong coupling constant, and $C_F=4/3$.

The main objective of this paper is to calculate the
single-transverse-spin dependent cross section $\Delta
\sigma(S_\perp)=[\sigma(S_\perp)-\sigma(-S_\perp)]/2$ coming from
the twist-three quark-gluon correlation in the polarized proton.
At large transverse momentum $P_{h\perp}\gg\Lambda_{\rm QCD}$, the
corresponding SSA is generated by the ETQS mechanism in terms of
twist-three transverse-spin dependent quark-gluon correlation
functions~\cite{qiu}. The difference between the physics of the
unpolarized cross section and transverse-spin dependent one is
that the latter involves an additional polarized gluon from the
polarized proton, which interacts with partons in the hard part,
in accordance with the twist-three nature of the observable. In
Fig.~1, we show a generic Feynman diagram for such a contribution.
The lower shaded oval of the diagram represents the
transverse-spin-dependent quark-gluon correlation function for the
polarized-proton target and is defined as \cite{qiu}:
\begin{eqnarray} \label{TF}
T_F(x_1,x_2) &\equiv &
\int\frac{d\zeta^-d\eta^-}{4\pi}e^{i(k_{q1}^+\eta^-+k_g^+\zeta^-)}
\, \epsilon_\perp^{\beta\alpha}S_{\perp\beta}
\label{eq1} \\
&&\left\langle PS|\overline\psi(0){\cal L}(0,\zeta^-)\gamma^+
gF_\alpha^{\ +}
(\zeta^-){\cal L}(\zeta^-,\eta^-)
\psi(\eta^-)|PS\right\rangle \nonumber \ ,
\end{eqnarray}
where the sums over color and spin indices are implicit,
$|PS\rangle$ denotes the proton state, $\psi$ the quark field, and
$F_{\alpha}^{\ +}$ the gluon field tensor. In Eq.~(\ref{eq1}),
$x_1=k_{q1}^+/P^+$ and $x_2=k_{q2}^+/P^+$ are the fractions of the
polarized proton's light-cone momentum carried by the initial quark lines in
Fig.~1, while $x_g=k_g^+/P^+ = x_2-x_1$ is the fractional momentum
carried by the gluon; ${\cal L}$ is the light-cone gauge link,
${\cal L}(\zeta_2,\zeta_1) = \exp\left(-ig\int^{\zeta_1}_{\zeta_2}
d\xi^- A^+(\xi^-)\right)$, that makes the correlation operator
gauge-invariant, and $\epsilon^{\alpha\beta}_\perp$ is the
2-dimensional Levi-Civita tensor with $\epsilon^{12}_\perp=1$.

\begin{figure}[t]
\begin{center}
\includegraphics[height=4.0cm]{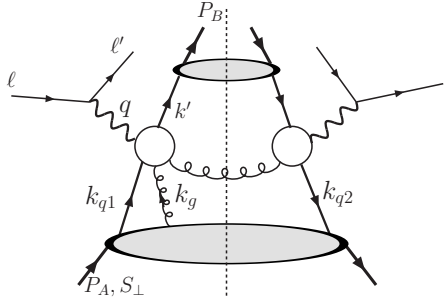}
\end{center}
\vskip -0.7cm \caption{\it A generic Feynman diagram contributing
to the Sivers-type single-transverse-spin asymmetry for inclusive
hadron production in semi-inclusive deep inelastic scattering.}
\end{figure}

\begin{figure}[t]
\begin{center}
\includegraphics[height=8.0cm]{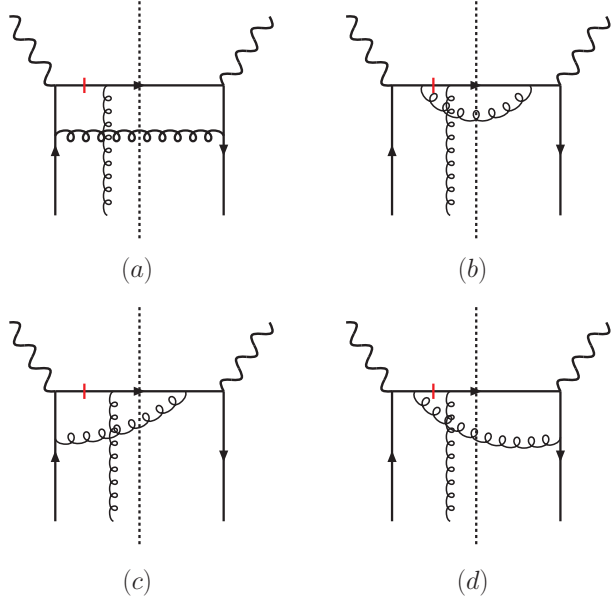}
\end{center}
\vskip -0.7cm \caption{\it Feynman diagrams making soft-pole
contributions to the single-transverse-spin dependent cross section.
The bars indicate the propagators where a soft pole arises.
The ``mirror'' diagrams for which the additional initial gluon attaches
on the right of the cut are not shown, but are included in the
calculations.}
\end{figure}

The strong interaction phase necessary for having a non-vanishing
SSA arises from the interference between an imaginary part of the
partonic scattering
amplitude with the extra gluon and the real scattering
amplitude without a gluon in Fig.~1.  The imaginary part is due to
the pole of the parton propagator associated with the integration
over the gluon momentum fraction $x_g$. Depending on which propagator's
pole contributes, $\Delta\sigma(S_\perp)$ may get contributions from
$x_g=0$ (``soft-pole'') \cite{qiu} and $x_g\neq 0$ (``hard-pole'')
\cite{JiQiuVogYua06,luo,guo}. When we calculate the partonic
scattering amplitudes, we have to attach the polarized gluon to
any propagator of the hard part represented by the light circles in
the diagram of Fig.~1. In particular, if the
polarized gluon attaches to the outgoing quark in the final state,
the on-shell propagation of the quark line will generate a soft
gluonic pole. A hard pole arises when internal quark propagators
go on-shell with nonzero $x_g$. In Figs.~2 and 3 we show the
relevant soft- and hard-pole partonic diagrams, respectively. There
are a total of eight diagrams contributing to the soft-pole part,
four of which we show in Fig.~2. The remaining four diagrams
can be obtained by attaching the gluon on the right side of the cut.
There are twelve diagrams for the hard-pole contributions, and again
only half of them are shown in Fig.~3. We note that only diagrams with an
$s$-channel quark propagator can have a hard pole. All diagrams in
Figs.~2 and 3 are crossed versions of the ones needed for the
SSA in the Drell-Yan process considered in \cite{JiQiuVogYua06}.

\begin{figure}[t]
\begin{center}
\includegraphics[height=8.0cm]{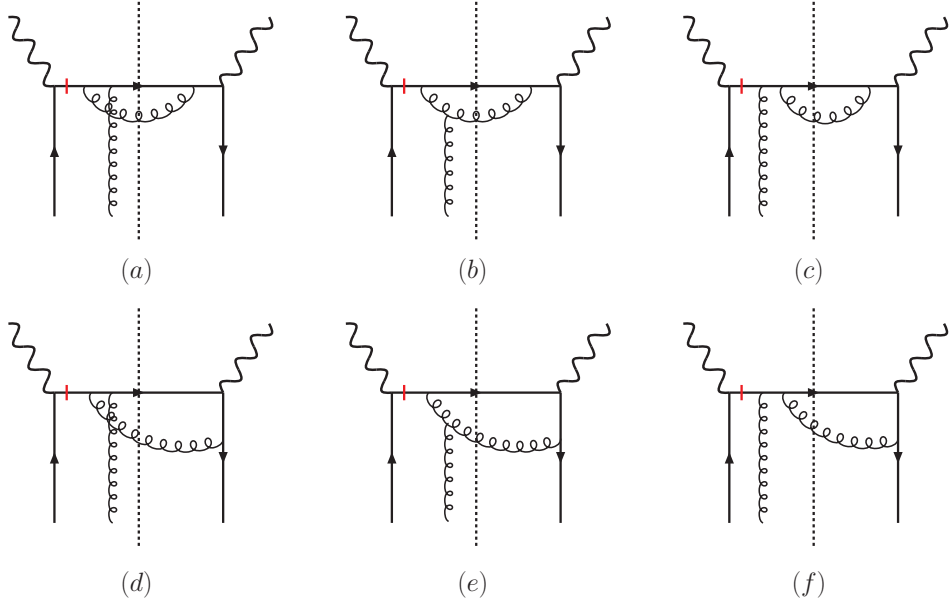}
\end{center}
\vskip -0.7cm \caption{\it Same as Fig.~2, but
for the hard-pole contributions.}
\end{figure}

The calculations of the soft-pole and hard-pole contributions
follow the same procedure as we used for the Drell-Yan process
\cite{JiQiuVogYua06}. We only give a brief outline here and refer
the reader to this reference for details. We perform our
calculations in a covariant gauge. The collinear expansion is the
central step in obtaining the final results. For example, in the
diagrams of Figs. 2 and 3, the dominant component of the momentum
of the polarized gluon is $x_gP+k_{g\perp}$. The contribution to
the single-transverse-spin asymmetry arises from terms linear in
$k_{g\perp}$ in the expansion of the partonic scattering
amplitudes. One important contribution to the $k_{g\perp}$
expansion comes from the on-shell condition for the outgoing
``unobserved'' gluon, whose momentum depends on $k_{g\perp}$. This
leads to a term involving the {\it derivative} of the correlation
function $T_F$. In addition, the soft and hard poles in the
diagrams may also arise as double poles \cite{qiu}, which will
lead to a derivative contribution as well. The hard-pole
contributions by the individual diagrams in Fig.~3 also give
derivative terms. However, the derivative contributions cancel out
in their sum, similar to what we found for the Drell-Yan case in
\cite{JiQiuVogYua06}. For example, the derivative contribution
from Fig.~3(a) is canceled out by part of 3(b), 3(c) by another
part of 3(b). The remaining contributions contain only
non-derivative terms. We note that in order to obtain the correct
result for the hard-pole contributions it is crucial to sum only
over physical polarization states of the ``unobserved'' gluon in
the Feynman diagrams.

Combining the contributions by all the diagrams, we find for the
single-transverse-spin dependent cross section:
\begin{eqnarray} \label{full}
    \frac{d\Delta\sigma(S_\perp)}{dx_Bdydz_hd^2\vec{P}_{h\perp}}\Big|_{{\cal
    V}_1}
      &=& - \frac{4\pi\alpha^2_{\rm em}S_{ep}}{z_h^3 Q^4}
      \epsilon^{\alpha\beta}S_\perp^\alpha P_{h\perp}^\beta
      \frac{\alpha_s}{2\pi^2}\int \frac{dxdz}{xz}\hat q
      (z)\delta\left(\vec{q}_\perp^{\; 2}-\frac{Q^2(1-\xi)(1-\hat \xi)}
{\xi\hat \xi}\right)\nonumber\\
      &&\times x_B\left(1-y+\frac{y^2}{2}\right)\left\{
\left(x\frac{\partial}{\partial x}T_F(x,x)\right)\left(\frac{1}{2N_c}\right)
      \frac{1-\xi}{\hat \xi \vec{q}_\perp^{\; 2}}~\hat
\sigma_{unp}\right.\nonumber\\
      &&+\left(-\frac{1}{2N_c}\right)
      T_F(x,x)\frac{\xi}{Q^2}\left[\frac{1+\hat
      \xi^2}{(1-\xi)^2(1-\hat \xi)^2}\right.\nonumber\\
      &&\left.+\frac{2\hat\xi(2-3\hat
      \xi)+(1-2\xi)(1+6\hat
      \xi^2-6\hat\xi)}{(1-\hat\xi)^2}\right]\nonumber\\
      &&\left.+T_F(x,x_B)\left(\frac{1}{2N_c}+C_F
      \hat \xi\right)\frac{\xi}{Q^2}\frac{1+\hat
      \xi^2\xi}{(1-\xi)^2(1-\hat\xi)^2}
\right\} \ ,
\end{eqnarray}
where $\hat\sigma_{unp}$ has been defined in Eq.~(\ref{sunp}). Again,
we have kept only the contribution associated with the tensor structure
${\cal V}_1$. All other terms have been neglected, because they
are suppressed by $q_\perp/Q$ in the limit of $q_\perp\ll Q$. Similar to
the Drell-Yan process, the hard part for the derivative term is
proportional to the unpolarized cross section. The last
term in the above equation comes from the hard-pole contributions.
As one can see, these are characterized by a dependence on the
quark-gluon correlation function $T_F(x,x_B)$, unlike the soft-pole
ones which enter with two identical momentum fractions in $T_F$.
We note in passing that we have also performed all calculations in a
frame where the initial proton and the produced hadron are collinear
and move in the $z$-direction. We found identical results for both the
soft-pole and the hard-pole contributions.

We point out that the derivative contribution in Eq.~(\ref{full}) agrees with
that derived in \cite{yujitalk}. Our non-derivative terms for the soft-pole
and the hard-pole contributions are new, however. We emphasize that even
though the derivative contribution is expected to dominate in some
kinematic situations~\cite{qiu}, the non-derivative parts become
of equal importance for $q_\perp\ll Q$, as we shall see shortly. Since it
is our goal in this paper to match the result obtained within the
ETQS formalism
to the one based on TMD factorization, it is crucial that we keep
the non-derivative parts. This is also to be seen in the context
that the bulk of the SIDIS event rate in experiment is generally
located at relatively modest $q_\perp$.

We also note that the angular correlation between the observed
hadron's transverse momentum $P_{h\perp}$ and the target proton's
polarization vector $S_\perp$ as shown in Eq.~(\ref{full}) is
characteristic of the contribution from the quark-gluon
correlation in the proton. Other contributions, like the
twist-three quark-gluon correlation in the fragmentation function,
will lead to a different angular correlation between these two
\cite{yujitalk}. With different angular dependence, these
contributions can be easily disentangled experimentally (see,
e.g.,\cite{dis}).

The results for the contributions by the structure function $W_1$
to the unpolarized and the single-transverse-spin dependent cross
sections given in Eqs.~(\ref{unp}) and~(\ref{full})
are valid 
when both $P_{h\perp}, Q \gg \Lambda_{\rm QCD}$.
In order to make contact with
the TMD factorization formalism, we shall now extrapolate our
results into the region of $\Lambda_{\rm QCD}\ll P_{h\perp}\ll Q$.
This is also the region exclusively dominated by the contributions
associated with the tensor ${\cal V}_1$ that we have considered.
In doing the expansion, we only keep the terms leading in
$P_{h\perp}/Q$, and neglect all higher powers. For
small $P_{h\perp}/Q$, the delta function in Eqs.~(\ref{unp})
and~(\ref{full}) can be expanded as \cite{meng}
\begin{equation}
\delta\left(\vec{q}_\perp^{\; 2}-\frac{Q^2(1-\xi)(1-\hat\xi)}{\xi\hat\xi}
\right)=\frac{\xi\hat
\xi}{Q^2}\left\{\frac{\delta(\xi-1)}{(1-\hat\xi)_+}+\frac{\delta(
\hat\xi-1)}{(1-\xi)_+}+
\delta(\xi-1)\delta(\hat\xi-1)\ln\frac{Q^2}{\vec{q}_\perp^{\; 2}}\right\} \ .
\end{equation}
Inserting this expression into Eq.~(\ref{unp}), we find for
the small-$P_{h\perp}$ behavior of the unpolarized
differential cross section \cite{meng}:
\begin{eqnarray}
    \frac{d\sigma}{dx_Bdydz_hd^2\vec{P}_{h\perp}}
      &=& \frac{4\pi\alpha^2_{\rm
      em}S_{ep}}{Q^4}\frac{\alpha_s}{2\pi^2}\frac{1}{\vec{P}_{h\perp}^2}
C_F\int \frac{dxdz}{xz}q(x)\hat q
      (z)\left\{\frac{1+\xi^2}{(1-\xi)_+}\delta(\hat\xi-1)\nonumber\right.\\
      &&\left.+\frac{1+\hat\xi^2}{(1-\hat\xi)_+}\delta(\xi-1)+
      \delta(\xi-1)\delta(\hat\xi-1)\ln\frac{z_h^2Q^2}{\vec{P}_{h\perp}^2}
\right\}\
      . \label{unx}
\end{eqnarray}
Similarly, for the single-transverse-spin dependent cross section,
we have
\begin{eqnarray}
    \frac{d\Delta\sigma(S_\perp)}{dx_Bdydz_hd^2\vec{P}_{h\perp}}
      &=& -\frac{4\pi\alpha^2_{\rm
      em}S_{ep}}{Q^4}\epsilon^{\alpha\beta}S_\perp^\alpha
      \frac{z_hP_{h\perp}^\beta}{(\vec{P}_{h\perp}^2)^2}
      \frac{\alpha_s}{2\pi^2}\int \frac{dxdz}{xz}\hat q
      (z)\nonumber\\
      &&\times \left\{\delta(\hat \xi-1)A+\delta(\xi-1)B\right\} \
      ,\label{plx}
\end{eqnarray}
where
\begin{eqnarray}
A&=& \frac{1}{2N_C} \left\{ \left[x\frac{\partial}{\partial
x}T_F(x,x)\right](1+\xi^2)
      +T_F(x,x-\widehat{x}_g)\frac{1+\xi}{(1-\xi)_+}\nonumber\right.\\
 &&\left.     +T_F(x,x)\frac{(1-\xi)^2(2\xi+1)-2}{(1-\xi)_+}\right\}
+C_F T_F(x,x-\widehat{x}_g)\frac{1+\xi}{(1-\xi)_+}\ ,
\label{qta}\\
B&=& C_F T_F(x,x)\left[\frac{1+\hat\xi^2}{(1-\hat\xi)_+}
+2\delta(\hat\xi-1)\ln\frac{z_h^2Q^2}{\vec{P}_{h\perp}^2}\right] \ ,
\label{qtb}
\end{eqnarray}
with $\widehat{x}_g\equiv(1-\xi)x=x-x_B$. We stress that both soft
poles and hard poles contribute to this result. Note that the
$T_F$ function for the hard-pole contribution reduces to
$T_F(x,x)$ at $\xi=1$. It turns out that this property is crucial
for obtaining the correct structure of the small-$P_{h\perp}$
limit of the cross section, consistent with the TMD factorization.
Because the contributions from all tensor structures other than
${\cal V}_1$ vanish in the limit of $P_{h\perp}\ll Q$, the above
results are the final results for the unpolarized and
(Sivers-type) single-transverse-spin dependent cross sections in
this kinematical regime.

Comparing the small-$P_{h\perp}$ behavior in Eqs.~(\ref{plx}), (\ref{qtb})
to the one we obtained for the Drell-Yan process at low pair transverse
momentum $q_\perp\ll Q$ \cite{JiQiuVogYua06}, we find that the hard
partonic parts are the same, with however an opposite sign.
This sign difference comes from the fact that in the Drell-Yan SSA
the strong interaction phase arises from initial-state
interactions, while in DIS it is due to final-state
interactions. Of course, the real physical asymmetries will also
depend on the size of the unpolarized quark distribution and fragmentation
functions and will not just differ by a sign. We note that this
universality (up to a sign) of the Drell-Yan and the SIDIS twist-three
partonic cross sections only happens at low transverse momentum.
At $q_\perp\sim Q$, there is no connection between the two processes
at all. The universality of the partonic hard parts at low transverse
momentum is actually a manifestation of the TMD factorization at
$P_{h\perp}\ll Q$, and of the universality of the TMD quark distributions
and fragmentation functions. We will discuss this in the
following section.

{\bf 3.} When $P_{h\perp}\ll Q$, we know that a
transverse-momentum-dependent factorization applies~\cite{JiMaYu04}.
Following this reference, the differential  SIDIS cross section may be
written as
\begin{eqnarray}
    \frac{d\sigma}{dx_Bdydz_hd^2\vec {P}_{h\perp}}
      &=& \sigma_0\times\left[F_{UU}^{(1)}+\sin(\phi_h-\phi_S)
|S_\perp| F_{UT}^{(1)}
      \right] \ ,
\end{eqnarray}
where $\sigma_0=4\pi\alpha^2_{\rm em}S_{ep}/{Q^4}\times
(1-y+y^2/2) x_B$, and where $\phi_{S}$ and $\phi_{h}$ are the
azimuthal angles of the proton's transverse polarization vector and of
the transverse momentum vector of the final-state hadron,
respectively. Again, we only keep the terms we are
interested in: $F_{UU}$ corresponds to the unpolarized cross section,
and $F_{UT}^{(1)}$ to the Sivers function contribution to the
single-transverse-spin asymmetry. Other contributions, for
example those related to the Collins effect~\cite{Col93}, may be incorporated
similarly~\cite{JiMaYu04}. $F_{UU}$ and $F_{UT}^{(1)}$ depend on the
kinematical variables, $x_B$, $z_h$, $Q^2$, $y$, and $P_{h\perp}$.
According to the TMD factorization formalism, they can be factorized
into TMD parton distributions and fragmentation functions, and soft and
hard parts. For example, $F_{UU}^{(1)}$ has the following
factorized form \cite{JiMaYu04}:
\begin{eqnarray} \label{FUU}
F_{UU}(x_B,z_h,Q^2,P_{h\perp})&=&\sum_{q=u,d,s,...} e_q^2\int
d^2\vec{k}_{\perp} d^2\vec {p}_{\perp}
      d^2\vec{\lambda}_\perp
   \nonumber \\
   && \times  q\left({x_B}, k_{\perp}\right)
    \hat q\left({z_h}, p_{\perp}\right)
    \left(S(\vec{\lambda}_\perp)\right)^{-1} \nonumber \\
&& \times H_{UU}\left(Q^2\right)
\delta^{(2)}\left(z_h\vec{k}_{\perp}+\vec{p}_{\perp}
+\vec{\lambda}_\perp- \vec{P}_{h\perp}\right) \ ,
\end{eqnarray}
where $q$ and $\hat q$ denote the unpolarized TMD quark distributions and
fragmentation functions, respectively. $H$ is a hard factor and is entirely
perturbative. It is a function of $Q\gg P_{h\perp}$ only. The
soft-factor $S$ is a vacuum matrix element of Wilson lines and
captures the effects of soft gluon radiation. Since the soft-gluon
contributions in the TMD distribution and fragmentation have not
been subtracted, the soft factor enters with inverse power.
We have not displayed the dependence of the TMD quark distribution
(fragmentation) functions on the variable $\zeta^2= (2v\cdot P)^2/v^2$
($\hat \zeta^2= (2\tilde v\cdot P_h)^2/\tilde v^2$), which serves
to regulate their light-cone singularities. Here, $v$ and $\tilde v$
are vectors off the light-cone. We finally introduce
the soft-gluon rapidity cut-off $\rho=\sqrt{(2v\cdot \tilde v)^2/
v^2\tilde v^2}$, on which the soft factor depends. In a special
coordinate frame, one may choose $x_B^2\zeta^2=\hat \zeta^2/z_h^2=\rho
Q^2$ \cite{JiMaYu04}. There is also explicit renormalization
scale dependence of the various factors in the factorization
formula which, too, has been omitted for simplicity.

Similarly to Eq.~(\ref{FUU}), the contribution to the
Sivers single-transverse-spin asymmetry can be factorized as
\begin{eqnarray} \label{FUT}
F_{UT}^{(1)}&=&\sum_{q=u,d,s,...} e_q^2\int d^2\vec{k}_{\perp}
d^2\vec {p}_{\perp}
      d^2\vec{\lambda}_\perp\frac{\vec{k}_\perp\cdot
\hat{\vec{P}}_{h\perp}}{M_P}
q_T\left({x_B}, k_{\perp}\right)
   \nonumber \\
&&\times  \hat q\left({z_h},p_{\perp}\right)
    \left(S(\vec{\lambda}_\perp)\right)^{-1}H_{UT}^{(1)}\left(Q^2\right)
    \delta^{(2)}\left(z_h\vec{k}_{\perp}+\vec{p}_{\perp} +
\vec{\lambda}_\perp-\vec{P}_{h\perp}\right) \ ,
\end{eqnarray}
where $\hat{\vec{P}}_{h\perp}$ is a unit vector in direction of
$\vec{P}_{h\perp}$ and $q_T$ is the Sivers TMD quark distribution.
The proton mass $M_P$ is used to normalize the Sivers function and
the unpolarized TMD quark distribution to the same mass dimension. For
the operator definition of the Sivers function, see for
example~\cite{JiQiuVogYua06}.

In order to make contact with the result for the ETQS formalism
of the previous section, we compute the various factors in
the factorization formulas~(\ref{FUU}), (\ref{FUT}) at large
transverse momentum ($P_{h\perp}\gg\Lambda_{\rm QCD}$), where their
dependence on $P_{h\perp}$ is perturbative. The unpolarized
quark distribution and fragmentation functions at large $P_{h\perp}$
can be expressed in terms of their respective $k_\perp$-integrated
distributions, multiplied by perturbatively calculable coefficients.
Their expressions are well known (see, for example, Ref.~\cite{JiMaYu04}).
For the quark distribution function, one has:
\begin{eqnarray}
q(x_B,k_\perp)&=&
\frac{\alpha_s}{2\pi^2}\frac{1}{\vec{k}_\perp^2}C_F\int\frac{dx}{x}
q(x) \left[\frac{1+\xi^2}{(1-\xi)_+}+\delta(\xi-1)
\left(\ln\frac{x_B^2\zeta^2}{\vec{k}_\perp^2}-1\right)\right]\ ,
\label{qktunp}
\end{eqnarray}
where $q(x)$ is the integrated quark distribution and $\xi=x_B/x$.
Likewise, the TMD quark fragmentation function is given by
\begin{eqnarray} \hat
q(z_h,p_\perp)&=&
\frac{\alpha_s}{2\pi^2}\frac{1}{\vec{p}_\perp^{\;2}}C_F\int\frac{dz}{z}
\hat q(z) \left[\frac{1+\hat\xi^2}{(1-\hat\xi)_+}+\delta(\hat
\xi-1)
\left(\ln\frac{\hat\zeta^2}{\vec{p}_\perp^2}-1\right)\right]\ ,
\label{ffun}
\end{eqnarray}
where $\hat q(z)$ is the integrated quark fragmentation function
and $\hat \xi=z_h/z$.

Similarly, the Sivers function at large $k_\perp$ can also be
calculated perturbatively. Because it is (naively)
time-reversal-odd, the only contribution comes from the
twist-three quark-gluon correlation function $T_F$ in Eq.~(\ref{TF}).
The calculation follows the same procedure as for our calculation
for the Drell-Yan process in \cite{JiQiuVogYua06}. The Feynman diagrams
are the same, the only difference being that the gauge-link propagators
have an opposite sign for their imaginary parts.
Carrying out the calculations accordingly, we find
\begin{eqnarray}
q_T(x_B,k_\perp)&=&-\frac{\alpha_s}{4\pi^2}\frac{2M_P}
{(\vec{k}_\perp^2)^2}\int\frac{dx}{x} \left\{A+C_FT_F(x,x)
\delta(\xi-1)\left(\ln\frac{x_B^2\zeta^2}{\vec{k}_\perp^2}-
1\right)\right\} \ , \label{sivpert}
\end{eqnarray}
where $A$ has been defined in Eq.~(\ref{qta}) and where
$\xi=x_B/x$. Indeed, as expected~\cite{BroHwaSch02,Col02,BelJiYua02},
we find that the Sivers function in DIS is the same as that in the
Drell-Yan process, but with an opposite sign. As is well-known
now~\cite{BroHwaSch02,Col02,BelJiYua02}, this sign difference
comes from the different directions of the gauge links for the
two processes: in DIS the gauge link arises from final-state interactions and
runs to positive light-cone infinity, while in Drell-Yan it is due to
initial-state interactions and goes to $-\infty$.

In order to calculate the explicit $P_{h\perp}$-dependence generated
by the TMD factorization, we let one of the transverse momenta
$\vec{k}_{\perp}$, $\vec{p}_\perp$, and $\vec{\lambda}_\perp$ be of the
order of $\vec{P}_{h\perp}$ and the others much smaller. When
$\vec{\lambda}_\perp$ is large, for example, we neglect $\vec{k}_\perp$
and $\vec{p}_\perp$ in the delta function, and the integrations over
these momenta yield either the ordinary quark distribution, or a
$k_\perp$ moment of the Sivers function. The latter is related to the
twist-three correlation \cite{BoeMulPij03}:
\begin{eqnarray}
\int d^2\vec{k}_\perp q(x,k_\perp) = q(x) \ ,
~~~~\int d^2\vec{k}_\perp \frac{\vec{k}_\perp^2}{M_P} q_T(k_\perp, x) =
-T_F(x,x)\ ,
\end{eqnarray}
where the minus sign on the right-hand-side of the second
equation is again due to the direction of the DIS gauge link. In case
$\vec{\lambda}_\perp$ is neglected in the delta function, one makes use
of the relation~\cite{JiMaYu04} $\int d^2\vec{\lambda}_\perp S(\lambda_\perp) = 1$.
We then obtain the following results for the unpolarized and single-transverse-spin
dependent cross sections:
\begin{eqnarray}
    \frac{d\sigma}{dx_Bdydz_hd^2\vec {P}_{h\perp}}
      &=& \sigma_0
      \frac{\alpha_s}{2\pi^2}C_F\frac{1}{\vec{P}_{h\perp}^2}\int\frac{dxdz}{xz}q(x)
      \hat q(z)\left\{\frac{1+\xi^2}{(1-\xi)_+}\delta(\hat \xi-1)\nonumber\right.\\
      &&\left.+
      \frac{1+\hat \xi^2}{(1-\hat
      \xi)_+}\delta(\xi-1)+\delta(\xi-1)\delta(\hat
      \xi-1)\ln\frac{Q^2z_h^2}{\vec{P}_{h\perp}^2}\right\} \ ,\\
\frac{d\Delta\sigma(S_\perp)}{dx_Bdydz_hd^2\vec {P}_{h\perp}}
      &=& \sigma_0
      \frac{\alpha_s}{2\pi^2}\epsilon_{\alpha\beta}S_\perp^\alpha
      \frac{-z_hP_{h\perp}^\beta}{(\vec{P}_{h\perp}^2)^2}\int\frac{dxdz}{xz}q(x)
      \hat
      q(z)\nonumber\\
      &&\times \left\{\delta(\hat \xi-1)A+\delta(\xi-1)B\right\}\
      ,
\end{eqnarray}
where $A$ and $B$ are defined as in Eqs.(\ref{qta}),~(\ref{qtb}). It is
evident that the above results reproduce the differential
cross sections in Eqs.(\ref{unx}),~(\ref{plx}).

{\bf 4.} In conclusion, we have demonstrated in this paper that the two
mechanisms for the Sivers-type single-transverse-spin asymmetry
in semi-inclusive deeply-inelastic scattering
are consistent at moderate transverse momentum, $\Lambda_{\rm QCD}\ll
P_{h\perp}\ll Q$. This provides an additional test of the unification
of the mechanisms discussed in~\cite{JiQiuVogYua06}. It will be important
to carry out a relevant experimental test of this unification.
Furthermore, our calculation also explicitly exemplifies the
process-dependence of the functions generating single-transverse-spin
asymmetries. We finally note that
another interesting SSA phenomenon in semi-inclusive DIS processes is associated with
the so-called Collins effect \cite{Col93}. A similar connection between
the twist-three quark-gluon correlation mechanism in fragmentation
\cite{Kanazawa:2000hz} and the Collins function should exist. An extension to
this case would be very interesting.

{\bf Acknowledgments.} X.~J. is supported by the U. S. Department
of Energy via grant DE-FG02-93ER-40762 and by a grant from Chinese
National Natural Science Foundation (CNSF). J.~Q. is supported in
part by the U. S. Department of Energy under grant No.
DE-FG02-87ER-40371. W.~V. and F.~Y. are finally grateful to RIKEN,
Brookhaven National Laboratory and the U.S. Department of Energy
(contract number DE-AC02-98CH10886) for providing the facilities
essential for the completion of their work.


\begin{thebibliography}
\frenchspacing
\bibitem{E704-Bunce}
  G.~Bunce {\it et al.},
  Phys.\ Rev.\ Lett.\  {\bf 36}, 1113 (1976); for later results on SSAs
in fixed-target hadronic scattering, see:
  D.~L.~Adams {\it et al.}  [E581 and E704 Collaborations],
  Phys.\ Lett.\ B {\bf 261}, 201 (1991);
 D.~L.~Adams {\it et al.}  [FNAL-E704 Collaboration],
  Phys.\ Lett.\ B {\bf 264}, 462 (1991);
K.~Krueger {\it et al.}, Phys.\ Lett.\ B {\bf 459}, 412 (1999).

\bibitem{dis}
A.~Bravar  [Spin Muon Collaboration],
Nucl.\ Phys.\ A {\bf 666}, 314 (2000);
A.~Airapetian {\it et al.}  [HERMES Collaboration],
Phys.\ Rev.\ Lett.\  {\bf 84}, 4047 (2000);
A.~Airapetian {\it et al.}  [HERMES Collaboration],
Phys.\ Rev.\ D {\bf 64}, 097101 (2001);
  Phys.\ Rev.\ Lett.\  {\bf 94}, 012002 (2005);
  M.~Diefenthaler [HERMES Collaboration],
AIP Conf.\ Proc.\  {\bf 792}, 933 (2005);
H.~Avakian  [CLAS Collaboration], talk presented at the RBRC
workshop ``Single-Spin Asymmetries'', Brookhaven National
Laboratory, Upton, New York, June 1-3, 2005, to appear in the
proceedings;
  V.~Y.~Alexakhin {\it et al.}  [COMPASS Collaboration],
  Phys.\ Rev.\ Lett.\  {\bf 94}, 202002 (2005).

\bibitem{rhic}
  J.~Adams {\it et al.}  [STAR Collaboration],
  Phys.\ Rev.\ Lett.\  {\bf 92}, 171801 (2004);
  S.~S.~Adler  [PHENIX Collaboration],
Phys.\ Rev.\ Lett.\  {\bf 95}, 202001 (2005);
  F.~Videbaek  [BRAHMS Collaboration],
AIP Conf.\ Proc.\  {\bf 792}, 993 (2005); arXiv:nucl-ex/0601008.

\bibitem{review} for reviews, see:
M.~Anselmino, A.~Efremov and E.~Leader,
Phys.\ Rept.\  {\bf 261}, 1 (1995) [Erratum-ibid.\  {\bf 281}, 399
(1997)];
  Z.~t.~Liang and C.~Boros,
  Int.\ J.\ Mod.\ Phys.\ A {\bf 15}, 927 (2000);
V.~Barone, A.~Drago and P.~G.~Ratcliffe,
Phys.\ Rept.\  {\bf 359}, 1 (2002).

\bibitem{Siv90}
D.~W.~Sivers,
Phys.\ Rev.\ D {\bf 41}, 83 (1990);
Phys.\ Rev.\ D {\bf 43}, 261 (1991).

\bibitem{Efremov}
  A.~V.~Efremov and O.~V.~Teryaev,
  Sov.\ J.\ Nucl.\ Phys.\  {\bf 36}, 140 (1982)
  [Yad.\ Fiz.\  {\bf 36}, 242 (1982)];
  A.~V.~Efremov and O.~V.~Teryaev,
  Phys.\ Lett.\ B {\bf 150}, 383 (1985).

\bibitem{qiu}
J.W.~Qiu and G.~Sterman,
Phys.\ Rev.\ Lett.\  {\bf 67}, 2264 (1991);
  Nucl.\ Phys.\ B {\bf 378}, 52 (1992);
Phys.\ Rev.\ D {\bf 59}, 014004 (1999).

\bibitem{BoeMulPij03}
D.~Boer, P.~J.~Mulders and F.~Pijlman,
Nucl.\ Phys.\ B {\bf 667}, 201 (2003).

\bibitem{MaWa03}
  J.~P.~Ma and Q.~Wang,
  Eur.\ Phys.\ J.\ C {\bf 37}, 293 (2004).

\bibitem{bacchetta}
  A.~Bacchetta,
  arXiv:hep-ph/0511085.

\bibitem{JiQiuVogYua06}
  X.~Ji, J.~W.~Qiu, W.~Vogelsang and F.~Yuan,
  arXiv:hep-ph/0602239;
  arXiv:hep-ph/0604023.

\bibitem{ColSop81}
J.~C.~Collins and D.~E.~Soper,
Nucl.\ Phys.\ B {\bf 193}, 381 (1981) [Erratum-ibid.\ B {\bf 213},
545 (1983)];
Nucl.\ Phys.\ B {\bf 197}, 446 (1982);
Nucl.\ Phys.\ B {\bf 194}, 445 (1982).

\bibitem{ColSopSte85}
J.~C.~Collins, D.~E.~Soper and G.~Sterman,
Nucl.\ Phys.\ B {\bf 250}, 199 (1985).

\bibitem{JiMaYu04}
  X.~Ji, J.~P.~Ma and F.~Yuan,
  Phys.\ Rev.\ D {\bf 71}, 034005 (2005);
Phys.\ Lett.\ B {\bf 597}, 299 (2004).

\bibitem{ColMet04}
J.~C.~Collins and A.~Metz,
Phys.\ Rev.\ Lett.\  {\bf 93}, 252001 (2004).

\bibitem{sidis}
  R.~b.~Meng, F.~I.~Olness and D.~E.~Soper,
  Nucl.\ Phys.\ B {\bf 371}, 79 (1992);
  Y.~Koike and J.~Nagashima,
  Nucl.\ Phys.\ B {\bf 660}, 269 (2003).


\bibitem{yujitalk} Y. Koike, talk given at the RIKEN/BNL Research
Center Workshop on ``Single-Spin Asymmetries", Brookhaven National
Laboratory, Upton, New York, June 1-3, 2005;
  H.~Eguchi, Y.~Koike and K.~Tanaka,
  arXiv:hep-ph/0604003.

\bibitem{Kanazawa:2000hz}
  Y.~Kanazawa and Y.~Koike,
  Phys.\ Lett.\ B {\bf 478}, 121 (2000);
  Phys.\ Rev.\ D {\bf 64}, 034019 (2001).

\bibitem{BroHwaSch02}
S.~J.~Brodsky, D.~S.~Hwang and I.~Schmidt,
Phys.\ Lett.\ B {\bf 530}, 99 (2002);
Nucl.\ Phys.\ B {\bf 642}, 344 (2002).

\bibitem{Col02}
J.~C.~Collins,
Phys.\ Lett.\ B {\bf 536}, 43 (2002).

\bibitem{BelJiYua02}
X.~Ji and F.~Yuan,
Phys.\ Lett.\ B {\bf 543}, 66 (2002);
A.~V.~Belitsky, X.~Ji and F.~Yuan,
Nucl.\ Phys.\ B {\bf 656}, 165 (2003).


\bibitem{Col93}
J.~C.~Collins,
Nucl.\ Phys.\ B {\bf 396}, 161 (1993).




\bibitem{luo}
  M.~Luo, J.~W.~Qiu and G.~Sterman,
  Phys.\ Rev.\ D {\bf 50}, 1951 (1994).

\bibitem{guo}
  X.~Guo,
  Phys.\ Rev.\ D {\bf 58}, 036001 (1998);
  Nucl.\ Phys.\ A {\bf 638}, 539C (1998).

\bibitem{meng}
  R.~Meng, F.~I.~Olness and D.~E.~Soper,
  Phys.\ Rev.\ D {\bf 54}, 1919 (1996).




\end{thebibliography}
\end{document}